**Anomalous Hall and Nernst effects in epitaxial films of topological kagome magnet $Fe_3Sn_2$**

Durga Khadka,[1,#] T. R. Thapaliya,[1,#] Sebastian Hurtado Parra,[2] Jiajia Wen,[3] Ryan Need,[4] James M. Kikkawa,[2] and S. X. Huang[1,*]

[1]Department of Physics, University of Miami, Coral Gables, Florida, 33146, USA

[2]Department of Physics and Astronomy, The University of Pennsylvania, Philadelphia, PA 19104, USA.

[3]Stanford Institute for Materials and Energy Sciences, SLAC National Accelerator Laboratory, Menlo Park, California 94025, USA

[4]Department of Materials Science and Engineering, University of Florida, Gainesville, Florida, 32611, USA

[#]Contributed equally.

[*]sxhuang@miami.edu

Abstract:

The topological kagome magnet (TKM) $Fe_3Sn_2$ exhibits unusual topological properties, flat electronic bands, and chiral spin textures, making it an exquisite materials platform to explore the interplay between topological band structure, strong electron correlations, and magnetism. Here we report the first synthesis of high-quality epitaxial (0001) $Fe_3Sn_2$ films with large intrinsic anomalous Hall effect close to that measured in bulk single crystals. In addition, we measured a large, anisotropic anomalous Nernst coefficient $S_{yx}$ of 1.26 μV/K, roughly 2-5x greater than that of common ferromagnets, suggesting the presence of Berry curvature sources near the Fermi level in this system. Crucially, the realization of high-quality $Fe_3Sn_2$ films opens the door to explore emergent interfacial physics and create novel spintronic devices based on TKMs by interfacing $Fe_3Sn_2$ with other quantum materials and by nanostructure patterning.

Topological semimetals (TSM) are three-dimensional (3D) analogs of graphene that possess crossing nodes (point or line) between the bulk conduction and valence bands at or near the Fermi energy [1,2]. These topologically protected, linearly dispersive band crossings (e.g., Weyl nodes) act as sources and sinks of Berry curvature in reciprocal ($k$) space and result in high carrier mobility. In addition, TSMs also exhibit exotic surface states such as Fermi arcs. Together, these features have attracted enormous attention over the past decade and make TSMs strong candidates for a host of applications, from spintronics to novel computing architectures [1-14].

Depending on the material's symmetries, TSMs can be divided into different classes: Weyl semimetals (WSM), Dirac semimetals (DSM), and nodal line semimetals (NLSM) [15]. Among TSMs, magnetic TSMs with broken time reversal symmetry are especially important and present a critical opportunity to study the interplay between topology and spin physics. Mounting theoretical evidence shows that band structure topology (e.g., number, location, and spacing of Weyl nodes) is intimately influenced by the underlying crystal and magnetic structures [16-20]. Magnetic microstructure, such as domain walls, are expected to further augment the band structure topology and give rise to unique surface states [2]. Moreover, the inherent spin-momentum locking of magnetic TSMs make them building blocks to realize other exotic quantum states such as quantum anomalous Hall and axion insulators [21].

Magnetic TSMs whose crystal structure contains a 2D kagome lattice (i.e., topological kagome magnets, TKM) have recently emerged as a topical research frontier [21-32]. In addition to Dirac or Weyl nodes, TKMs may host flat bands with zero dispersion and strong electron correlations. These flat bands are of fundamental interest and can support interaction-driven ferromagnetism, chiral superconductivity, Wigner crystals, and fractional quantum Hall states [33]. Flat bands are also posited to play a crucial role [34] in the superconductivity of twisted bilayer graphene [35]. Recently, a number of metallic TKMs ($Mn_3Sn$, $Fe_3Sn_2$, $FeSn$, $Co_3Sn_2S_2$, etc.) have been reported to exhibit unprecedented properties – such as the coexistence of Dirac fermions and flat bands, and negative flat band magnetism [21-31,36]. The intense studies of TKMs have focused primarily on bulk crystals due to the limited availability of high-quality thin films. However, thin films offer several advantageous tuning knobs not available in bulk samples, including: strengthening electron correlations and enhancing the role of topological surface states through reduced dimensionality, creating controlled interfaces between TKMs and other quantum materials which may host emergent phenomena, and enabling technological device fabrication based upon the novel physics of TKMs. In this work, we report the first fabrication of high-quality epitaxial (0001) $Fe_3Sn_2$ films that exhibit a large anomalous Hall effect (AHE) as observed in bulk single crystals. Furthermore, for the first time, we report large anomalous Nernst effects (ANE) in this system for both in-plane and out-of-plane temperature gradients.

$Fe_3Sn_2$ has a layered rhombohedral crystal structure (space group $R\bar{3}m$, *Fig.1a*) with lattice constants $a = 5.338$ Å and large $c = 19.789$ Å [26]. It has complex stacking of Sn layers (*Fig.1b*) and $Fe_3Sn$ layers (*Fig.1c*). In the $Fe_3Sn$ layers (*Fig.1c*), Fe atoms form a ferromagnetic kagome lattice with corner-sharing triangles and Sn atoms located at the center of the hexagons. Two adjacent $Fe_3Sn$ layers with different in-plane lattice offsets are separated by a honeycomb Sn layer (*Fig.1b*). $Fe_3Sn_2$ has a high Curie temperature ($T_C$) around 670 K and a saturation magnetization of 1.9 $\mu_B$/Fe (650 *emu/cc*) at low temperature [26]. Several pioneering studies on bulk Fe$_3$Sn$_2$ have shown evidence of topological band structure and properties, including: 1) Weyl nodes at the Fermi energy ($E_F$) which can be moved in reciprocal space by rotating the magnetization direction [37]; 2) quasi-2D Dirac cones (about 70 meV below $E_F$) with a mass gap of 30 meV [26]; 3) magnetization-driven giant nematic energy shift [27]; and 4) flat bands near the Fermi energy [24]. Furthermore, Fe$_3$Sn$_2$ exhibits a large anomalous Hall effect (AHE) [26,38] and magnetic skyrmions (topological spin textures in real space) [39]. Note that the large AHE has a substantial intrinsic contribution from Berry curvature[26].

We grew thin films of Fe$_3$Sn$_2$ by co-sputtering Fe and Sn targets at substrate temperatures of $T_S \approx 400\,°C$ in a high vacuum magnetron sputtering system with base pressure better than $5 \times 10^{-8}$ torr. The stoichiometry of Fe$_3$Sn$_2$ was evaluated from the sputtering rates of Fe and Sn and confirmed by x-ray diffraction (XRD) and magnetization measurements. Prior to Fe$_3$Sn$_2$ deposition, epitaxial seed layers of first (111) Pt (~ 3 nm) and then (0001) Ru (~ 3 nm) were deposited on (0001) Al$_2$O$_3$ substrates. The excellent quality of these seed layers is demonstrated by the high intensities and Laue oscillations of the Pt (111) and Ru (0001) peaks in XRD (*Fig.2a*). Ru has a hexagonal structure with lattice constant $a$ = 2.706 Å (2$a$ = 5.412 Å). The lattice mismatch between Ru and bulk Fe$_3$Sn$_2$ ($a_{bulk}$ = 5.338 Å) is only ~ 1.4%, which facilitates the high-quality epitaxial growth of Fe$_3$Sn$_2$. As shown in *Fig.2a*, only the (000*l*) ($l$ = 3, 6, 9) peaks of Fe$_3$Sn$_2$ are observed in the coupled XRD scan (2θ-ω scan) without any noticeable impurity peaks. Lattice constants of the film were calculated to be $c_{film}$ = 19.81 Å and $a_{film}$ = 5.38 Å. Note that $a_{film}$ is about 0.4% less than 2$a_{Ru}$ and 0.8% larger than $a_{bulk}$, indicating that the films are strained. Laue oscillations from the Fe$_3$Sn$_2$ (0009) peak can be seen in inset of *Fig.2a*, superposed on 3 nm Pt (111) peak. The oscillations have period of 0.31° corresponding to a thickness of 28.5 nm. Importantly, high resolution XRD (HR-XRD) rocking curves, collected with monochromating incident optics and a triple axis analyzer, show a very sharp peak with full width half maximum (FWHM) of 0.002° (*Fig.2b*) indicating the film has exquisite mosaic, nearly matching that of the Al$_2$O$_3$ substrate. The double axis HR-XRD rocking curve (inset of *Fig.2b*), whose breadth includes contributions of mosaic (0.002°) and *d*-spacing spread, shows a sharp peak (FWHM of 0.07°) sitting on top of a weak broad peak with FWHM of 0.5°. The *d*-spacing spread is likely due to relaxation of strain and/or small compositional variations from co-sputtering. The epitaxial relationship (Al$_2$O$_3$[10$\bar{1}$0] || Pt[110] || Ru[2$\bar{1}\bar{1}$0] || Fe$_3$Sn$_2$[2$\bar{1}\bar{1}$0]) between the substrate, seed layers, and Fe$_3$Sn$_2$ is confirmed by the diffraction peaks of in-plane φ scans (*Fig.2c*) at (11$\bar{2}$6) Al$_2$O$_3$, (002) Pt, (11$\bar{2}$2) Ru, and (20$\bar{2}$5) Fe$_3$Sn$_2$ planes, respectively. Moreover, the FWHM of φ scan (*Fig.2d*) of Fe$_3$Sn$_2$ is only 0.16° which indicates nearly perfect in-plane alignments of crystalline axes between different grains. The XRD reflectivity measurement (inset of *Fig.2d*) shows clear oscillations from low angles to above 6°, indicating flat surface. These XRD measurements conclusively demonstrate the high-quality epitaxial growth of (0001) Fe$_3$Sn$_2$ films.

In the bulk, Fe$_3$Sn$_2$ is reported to be a non-collinear frustrated ferromagnet with zero remanence that exhibits interesting spin textures such as magnetic skyrmion and spin glass phases [39]. At $T$ = 300 K, the magnetic easy axis is along the *c*-axis with saturation field $H_S$ of 7 kOe ($H$ || *c*-axis) and 4 kOe ($H$ || *ab*-plane) [39]. As temperature decreases, the system undergoes reorientation of the magnetic easy axis from the *c*-axis to *ab* plane resulting in $H_S$ of about 12 kOe ($H \parallel c$-axis) and 2 kOe ($H \parallel ab$-plane) at $T$ = 6 K [39]. In contrast, in a 60 nm (0001) Fe$_3$Sn$_2$ epitaxial film, we observed nearly square hysteresis loops for $H \parallel ab$-plane (*Fig.3a*) with $M_S$ of 610 emu/cc and relative remanence close to unity. Due to the shape anisotropy ($4\pi M_S \sim 8$ kOe), the $H_S$ is greatly reduced relative to the bulk values, and only reaches roughly 100 Oe at $T$ = 300 K and 180 Oe at $T$ = 10 K. The much smaller in-plane switching field and nearly full remanence is useful for spintronic applications and to study the magnetization induced variations of topological properties such as the aforementioned nematic energy shift of electronic state induced by rotating magnetization in the *ab*-plane [27].

Bulk Fe$_3$Sn$_2$ is known to exhibit a large anomalous Hall effect (AHE) [26]. For $T > 100$ K, the AHE in Fe$_3$Sn$_2$ single crystals ($H \parallel c$-axis) is dominated by the intrinsic contribution arising from the Berry curvature generated by the massive Dirac or Weyl fermions [26,40]. The intrinsic anomalous Hall conductivity ($\sigma_{AHE}$) is around 170 $\Omega^{-1}$cm$^{-1}$ and depends only weakly on the temperature [26]. *Fig.3b* shows field dependent Hall Resistivity ($\rho_H$, sum of anomalous Hall $\rho_{AHE}$ and ordinary Hall $R_0 H$, the contribution from buffer layer is not subtracted) of a Pt(2.5nm)/Ru(2nm)/Fe$_3$Sn$_2$(80 nm) film with a patterned 6-terminal Hall bar structure (width of 20 μm). In contrast to the in-plane magnetization loops with clear hysteresis, field dependent $\rho_H$ ($H \parallel c$-axis) shows negligible hysteresis or remanence (*Fig.3b*). Instead, $\rho_H$ increases nearly linearly from $H$ = 0 to $H = H_1$ (~ 5 kOe, $T$ = 300 K). From $H_1$ to $H_S$ (~14 kOe), $d\rho_H/dH$ decreases and shows a hump-like feature (inset of *Fig.3b*) which may reflect the rich domain/spin textures. Note this hump-like feature is also observed in the susceptibility measurements ($\chi$ vs $H$) in bulk Fe$_3$Sn$_2$ and is an indication of magnetic bubble or skyrmion phases [39]. Above $H_S$, $\rho_H$ reaches a saturation value ($\rho_H)_S$ of 4.8 μΩ·cm ($T$ = 300 K) with small linear background from ordinary Hall effect ($R_0 H \ll \rho_{AHE}$). At $T$ = 10 K, similar field dependent $\rho_H$ loop is observed with $H_S$ about 18 kOe but much less ($\rho_H)_S$ of 1.2 μΩ·cm.

The film resistivity ($\rho_{xx}$, including the contribution from buffer layer Pt/Ru) is about 170 μΩ·cm at $T$ = 300 K. Metallic behavior is observed over the entire temperature range (10 K – 300 K) and $\rho_{xx}$ decreases nearly linearly from $T$ = 300 K to $T$ = 50 K (left of *Fig.3c*). At low temperatures, $\rho_{xx}$ reaches a residue value of 86 μΩ·cm and shows no evidence of carrier localization. To obtain (Hall) resistivity of Fe$_3$Sn$_2$, we fabricated a separate buffer layer Pt(2.5nm)/Ru(2nm) and measured its temperature dependent resistivity $\rho_{xx}^b$. For the buffer/Fe$_3$Sn$_2$ film, the buffer layer is electrically connected in parallel with Fe$_3$Sn$_2$. The resistivity of Fe$_3$Sn$_2$ $\rho_{xx}^{\text{Fe-Sn}}$ can then be calculated by

$$\rho_{xx}^{\text{Fe-Sn}} = \frac{t_{\text{Fe-Sn}}}{t/\rho_{xx} - t_b/\rho_{xx}^b}, \quad (1)$$

where $t_{\text{Fe-Sn}}$ (80 nm) and $t_b$ (4.5 nm) are the thicknesses of Fe$_3$Sn$_2$ and buffer layer, respectively. At $T$ = 300 K, $\rho_{xx}^{\text{Fe-Sn}}$ is about 202 μΩ·cm (bottom inset of *Fig.3c*), which is close to the value in bulk single crystal (e.g., ~190 μΩ·cm in [26]).

To obtain temperature dependent $\rho_{AHE}$, we measured $\rho_H$ at +/- 30 kOe and +/- 40 kOe to remove the offset of longitudinal resistance and the contribution of ordinary Hall resistivity. As shown in the right *y*-axis of *Fig.3c*, $\rho_{AHE}$ increases about 4 times from 1.26 μΩ·cm at 10 K to 4.78

μΩ·cm at 300 K. We now discuss how to subtract the contribution of the buffer layer which does not generate anomalous Hall voltage. The measured Hall voltage $V_H$ is partial of anomalous Hall voltage ($V_H^{Fe-Sn}$) generated by Fe$_3$Sn$_2$ (top inset of *Fig.3c*), which can be derived as

$$V_H^{Fe-Sn} = V_H \times \left(1 + \frac{\rho_{xx}^{Fe-Sn} \times t_b}{\rho_{xx}^b \times t_{Fe-Sn}}\right). \quad (2)$$

Therefore, the anomalous Hall resistivity of Fe$_3$Sn$_2$ is

$$\rho_{AHR}^{Fe-Sn} = \rho_{AHR} \times \frac{\rho_{xx}^{Fe-Sn}}{\rho_{xx}} \times \left(1 + \frac{\rho_{xx}^{Fe-Sn} \times t_b}{\rho_{xx}^b \times t_{Fe-Sn}}\right). \quad (3)$$

$\rho_{AHR}^{Fe-Sn}$ is about 1.72 μΩ·cm at 10 K and 7.17 μΩ·cm at 300 K (bottom inset of *Fig.3c*).

With buffer contribution subtracted from the measured values, we plot $\rho_{AHR}^{Fe-Sn}$ as a function of $(\rho_{xx}^{Fe-Sn})^2 M_S$ (*Fig.3d*). Clearly, the curve follows a straight line which intercepts the origin. This is a very strong evidence showing that the AHE in the Fe$_3$Sn$_2$ film originates from the intrinsic contribution of Berry curvature [41] without notable extrinsic contribution. Consequently, the intrinsic anomalous Hall conductivity ($|\sigma_{AHE}| \approx (\rho_{AHE})/(\rho_{xx})^2$, note that the magnetoresistance is small in the field and temperature range) is nearly a constant of 176 Ω$^{-1}$cm$^{-1}$ (corresponding to 0.3 $e^2/h$ per kagome bilayer) from 10 K to 300 K, which is close to the value from the bulk intrinsic contribution.

Like the AHE, the anomalous Nernst effect (ANE) is another transport property bearing signature of Berry curvature. While the intrinsic AHE results from the integration of Berry curvature over the Brillouin zone, the ANE is determined by the Berry curvature at the Fermi energy [42,43]. To our knowledge, no ANE measurements have been reported in bulk Fe$_3$Sn$_2$. In the ANE measurements, instead of applying electric current as in AHE measurements, a thermal current (temperature gradient $\nabla T$) is applied. Here we show large ANE in (0001) Fe$_3$Sn$_2$ films for both $\nabla T \parallel ab$-plane and $\nabla T \parallel c$-axis.

In the Nernst effect measurements, magnetic field (magnetization *M*), temperature gradient ($\nabla T$), and transverse voltage (Nernst electric field *E*) are mutually perpendicular to each other (*Fig.4a*), giving $E = Q_S \nabla T \times (4\pi M) = S_{yx} \nabla T \times m$ (left hand side of *Fig.4a*), where $Q_S$ is the anomalous Nernst coefficient, $S_{yx}$ is the transverse Seebeck coefficient ($S_{yx} = E_{yx}/\nabla T_x$), and *m* is the unit vector of magnetization [42]. $S_{yx} = \theta_{ANE} S_{xx}$, where $S_{xx}$ is the Seebeck coefficient and $\theta_{ANE}$ is the anomalous Nernst angle. *Fig.4b* shows $E_{yx}$ ($\nabla T \parallel ab$-plane, $\nabla T = 2$ K/mm) as a function of field, which resembles the AHE curve shown in *Fig.3c*. The inset of *Fig.4b* shows the linear relationship between $E_{yx}$ ($\nabla T_x$) and the applied heating power. The measured $S_{yx}$ is about 1 μV/K. Note that Equ.2 can be applied here, giving ($S_{yx}$)$_{Fe-Sn}$ of 1.26 μV/K after subtracting buffer contribution, which is noticeably larger than the values of conventional ferromagnets (e.g., 0.2-0.3 μV/K for Co/Fe [42], 0.46 μV/K for Py shown in *Fig.4c*). The measured $S_{xx}$ is -14 μV/K, resulting a high $\theta_{ANE}$ of -7%.

*Fig.4d* shows Nernst electric field $E_{yz}$ as a function of field for the out-of-plane temperature gradient ($\nabla T_Z$). $E_{yz}$ shows a clear hysteresis loop which resembles the magnetization loop in *Fig.3a*. $E_{yz}$ is as high as 2.5 μV/mm. In this geometry (right hand side of *Fig.4a*), ($\nabla T_Z$)$_{film}$ is practically impossible to be measured by usual means due to the small thickness (order of 10 nm). To estimate $(S_{yz})_{Fe-Sn}$ for $H \parallel ab$-plane, we measure ANE in Py films (also on $Al_2O_3$ substrate with the same heating power) and compare the Nernst signals. $(E_{yz})_{Py}$ is about 0.62 µV/mm (*Fig.4e*), which is 4 time less than that of $Fe_3Sn_2$ films. Assuming heat current flows uniformly from films to substrate, $(\nabla T_Z)_{Fe-Sn} \times \kappa_{Fe-Sn} \approx (\nabla T_Z)_{substrate} \times \kappa_{substrate} \approx (\nabla T_Z)_{Py} \times \kappa_{Py}$, where $\kappa$ is the thermal conductivity. Thus $(S_{yz})_{Fe-Sn} = [(E_{yz})_{Fe-Sn} / (E_{yz})_{Py}] \times (\kappa_{Fe-Sn} / \kappa_{Py}) \times (S_{yz})_{Py} \approx 1.85 \times (\kappa_{Fe-Sn} / \kappa_{Py})$ µV/K. $\kappa_{Py}$ is about 30 W/m-K while $\kappa_{Fe-Sn}$ has not been reported yet. As an estimation, assuming Wiedemann–Franz law applies and $\rho_c \approx \rho_{ab} = 170$ µΩ·cm in $Fe_3Sn_2$, $\kappa_{Fe-Sn} = L_0 \times T(= 300 \text{ K})/\rho_{ab} \approx 4.3$ W/m-K, where $L_0$ is the Lorentz number. Therefore, $(S_{yz})_{Fe-Sn}$ for $H \parallel ab$-plane is around 0.27 µV/K (0.34 µV/K if buffer contribution is subtracted), which is only a quarter of $S_{yx}$ (1 µV/K). These results suggest a possible large anisotropy of ANE, due to the intrinsic contribution of Berry curvature from the 2D kagome plane as suggested in the anisotropic AHE results on $Fe_3Sn_2$ single crystals [26]. Our results highlight the importance of performing thorough studies of thermoelectric properties in $Fe_3Sn_2$ single crystals, which can provide further insight into the topological band structure of $Fe_3Sn_2$.

In summary, we report the first fabrication of epitaxial (0001) $Fe_3Sn_2$ films with high crystalline quality as evidenced by small mosaicity (rocking curve FWHM = 0.002º), high in-plane domain alignment (in-plane φ scan FWHM = 0.16º), and flat interfaces/surface (Laue oscillations and reflectivity) critical for the patterning of nanostructured devices. The films possess distinct magnetic properties of the bulk samples, in particular, they show reduced in-plane coercive fields but greater relative remnant magnetization. A large intrinsic AHE is observed in our films with a Hall conductivity close to the intrinsic value measured on bulk crystals, suggesting that our thin, strained films retain Berry curvature source/sinks in the band structure. We also report, for the first time, possibly large anisotropic ANE (up to ~ 1.26 µV/K) that also points towards Berry curvature features intrinsic to topological band structures. These ANE results call for further investigation of thermoelectric properties of $Fe_3Sn_2$. Moreover, the availability of high-quality $Fe_3Sn_2$ thin films opens up the possibility to tune properties of $Fe_3Sn_2$ through strain and interfacial engineering.

S.H.P. and J.M.K acknowledge the support by the Penn NSF MRSEC DMR-1720530. J.W. acknowledges the support by the Department of Energy, Office of Science, Basic Energy Sciences, Materials Sciences and Engineering Division, under contract DE-AC02-76SF00515.

Figure captions:

Fig.1: (a) Crystal structure of $Fe_3Sn_2$ with stacking of $Fe_3Sn$ and Sn layers. (b) Sn layer with honeycomb lattice. (c) $Fe_3Sn$ layer with Fe kagome lattice.

Fig.2: (a) Representative XRD $2\theta$ scan of a 60 nm (0001) $Fe_3Sn_2$ film. Inset of (a): XRD $2\theta$ scan of a 28.5 nm $Fe_3Sn_2$ film showing (0009) peak accompanied by clear Laue oscillations. (b) Triple axis HR-XRD $\omega$ scan (rocking curve) of $Fe_3Sn_2$ (0009) peak in the inset of (a). Inset of (b): Double axis HR-XRD $\omega$ scan. (c) In-plane $\varphi$ scans of $Fe_3Sn_2$, Ru, Pt, and $Al_2O_3$. (d) Fine $\varphi$ scan of $Fe_3Sn_2$. Inset of (d): XRD reflectivity scan.

Fig.3: (a) Magnetic hysteresis loop (field in the *ab*-plane) for a 60 nm $Fe_3Sn_2$ film at $T = 300$ K (blue squares), and $T = 10$ K (red squares), respectively. (b) Hall resistivity as a function of field for a Pt(2.5nm)/Ru(2nm)/$Fe_3Sn_2$(80nm) film at $T = 300$ K (blue squares), and $T = 10$ K (red squares), respectively. Inset: First derivative of Hall resistivity as a function of field. (c) Resistivity (left) and anomalous Hall resistivity (right) as a function of temperature for the Pt(2.5nm)/Ru(2nm)/$Fe_3Sn_2$(80nm) film, and for $Fe_3Sn_2$ with buffer contribution subtracted (bottom inset). Top inset: Schematic circuit diagram showing contribution from buffer layer (Pt/Ru). (d) Anomalous Hall resistivity of $Fe_3Sn_2$ with buffer contribution subtracted as a function of $(\rho_{xx})^2 M_S$, where $\rho_{xx}$ ($M_S$) is the resistivity (saturated magnetization) of $Fe_3Sn_2$. The dashed linear lines is guide to eyes and intercepts to the origin. Inset: Anomalous Hall conductivity of $Fe_3Sn_2$ as a function of $(\sigma_{xx})^2 M_S$.

Fig.4: (a) Schematic diagrams of ANE measurements for in-plane (left) and out-of-plane (right) temperature gradient. (b), (c) Nernst electric field $E_{yx}$ ($\nabla T_x$, $H \parallel c$-axis) as a function of field for a Pt(2.5nm)/Ru(2nm)/$Fe_3Sn_2$(80nm) film (b) and a 40 nm Py (c) film. Inset of (b): $E_{yx}$ (left *y*-axis) and $\nabla T_x$ (right *y*-axis) as a function of heating power. (d), (e) $E_{yz}$ ($\nabla T_z$, $H \parallel ab$-plane) as a function of field for $Fe_3Sn_2$ (d) and Py (e) films.

Figure 1:

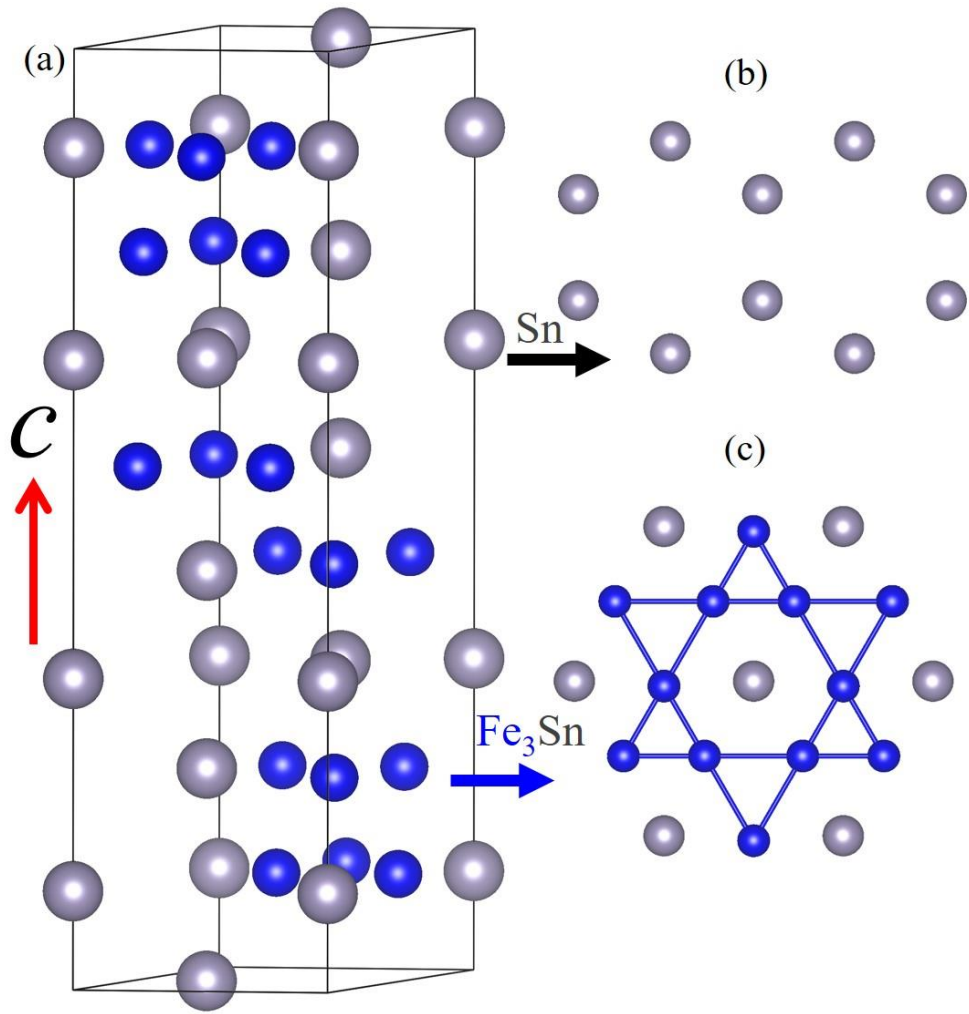

Fig.1: (a) Crystal structure of $Fe_3Sn_2$ with stacking of $Fe_3Sn$ and Sn layers. (b) Sn layer with honeycomb lattice. (c) $Fe_3Sn$ layer with Fe kagome lattice.

Figure 2:

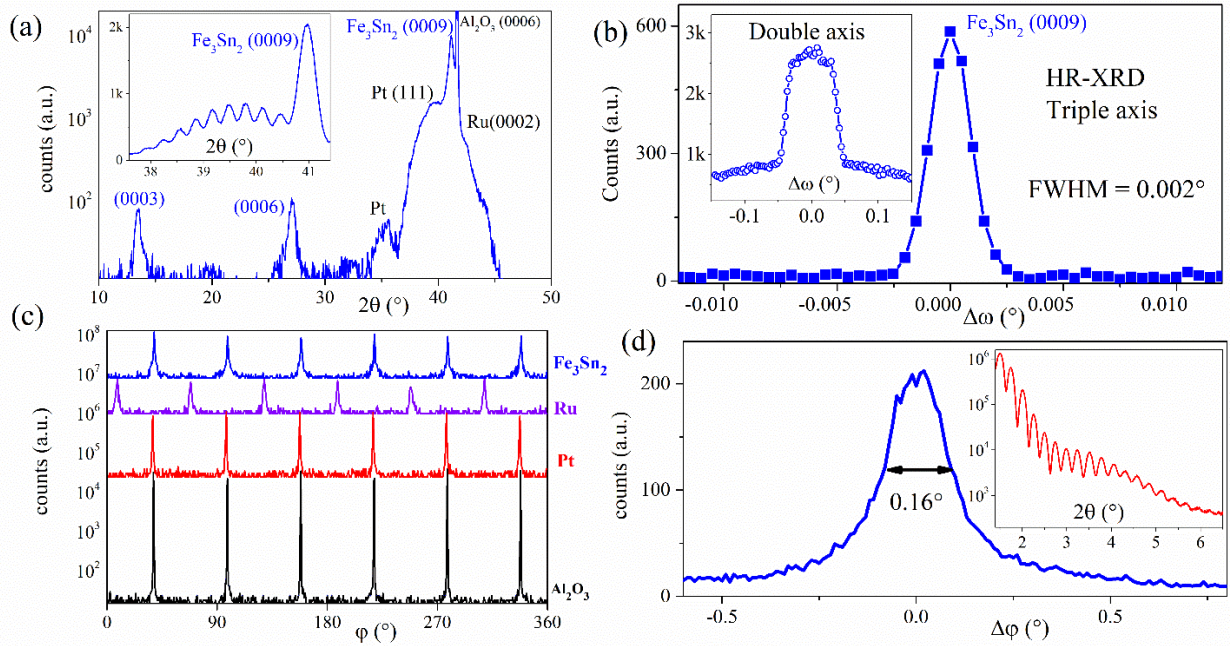

Fig.2: (a) Representative XRD 2θ scan of a 60 nm (0001) Fe$_3$Sn$_2$ film. Inset of (a): XRD 2θ scan of a 28.5 nm Fe$_3$Sn$_2$ film showing (0009) peak accompanied by clear Laue oscillations. (b) Triple axis HR-XRD ω scan (rocking curve) of Fe$_3$Sn$_2$ (0009) peak in the inset of (a). Inset of (b): Double axis HR-XRD ω scan. (c) In-plane φ scans of Fe$_3$Sn$_2$, Ru, Pt, and Al$_2$O$_3$. (d) Fine φ scan of Fe$_3$Sn$_2$. Inset of (d): XRD reflectivity scan.

Figure 3:

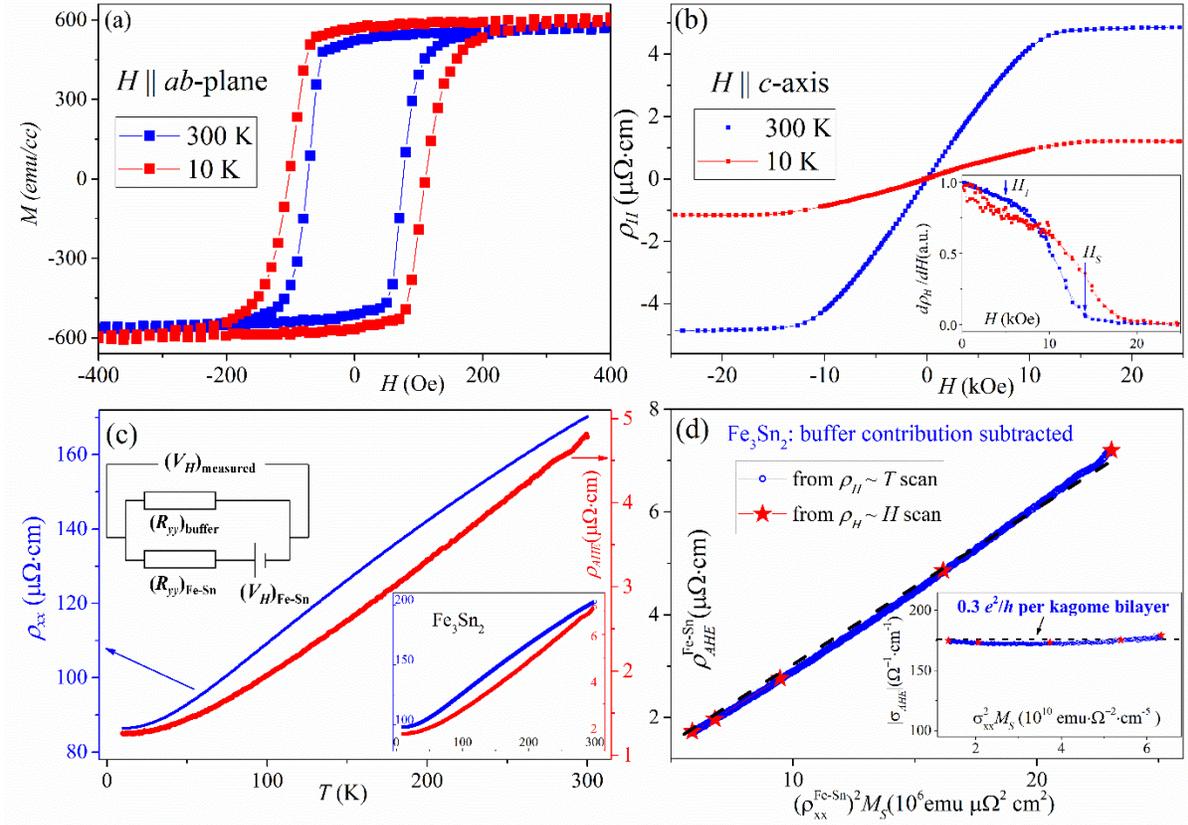

Fig.3: (a) Magnetic hysteresis loop (field in the *ab*-plane) for a 60 nm $Fe_3Sn_2$ film at $T = 300$ K (blue squares), and $T = 10$ K (red squares), respectively. (b) Hall resistivity as a function of field for a Pt(2.5nm)/Ru(2nm)/$Fe_3Sn_2$(80nm) film at $T = 300$ K (blue squares), and $T = 10$ K (red squares), respectively. Inset: First derivative of Hall resistivity as a function of field. (c) Resistivity (left) and anomalous Hall resistivity (right) as a function of temperature for the Pt(2.5nm)/Ru(2nm)/$Fe_3Sn_2$(80nm) film, and for $Fe_3Sn_2$ with buffer contribution subtracted (bottom inset). Top inset: Schematic circuit diagram showing contribution from buffer layer (Pt/Ru). (d) Anomalous Hall resistivity of $Fe_3Sn_2$ with buffer contribution subtracted as a function of $(\rho_{xx})^2 M_S$, where $\rho_{xx}$ ($M_S$) is the resistivity (saturated magnetization) of $Fe_3Sn_2$. The dashed linear lines is guide to eyes and intercepts to the origin. Inset: Anomalous Hall conductivity of $Fe_3Sn_2$ as a function of $(\sigma_{xx})^2 M_S$.

Figure 4:

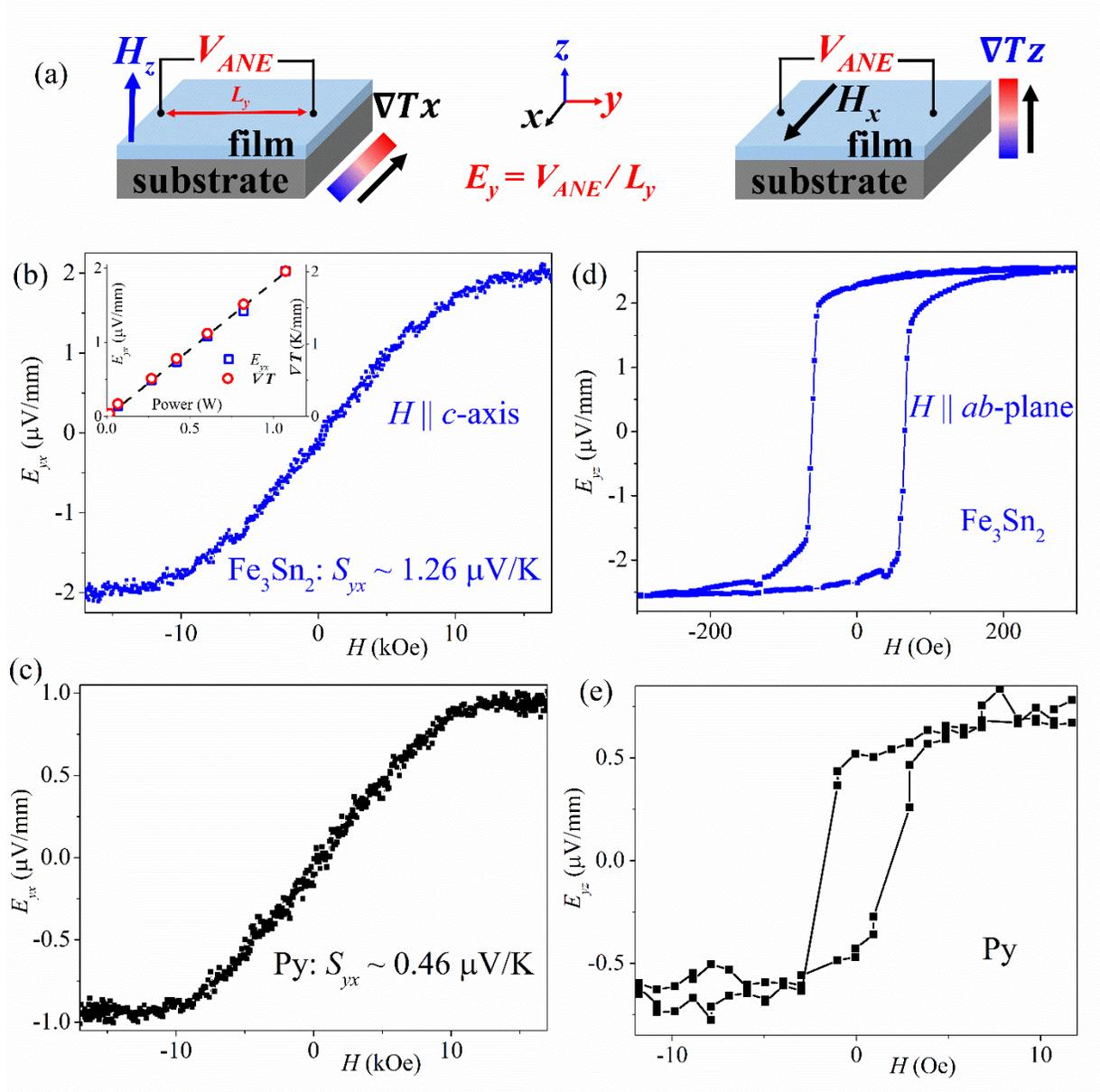

Fig.4: (a) Schematic diagrams of ANE measurements for in-plane (left) and out-of-plane (right) temperature gradient. (b), (c) Nernst electric field $E_{yx}$ ($\nabla T_x$, $H \parallel c$-axis) as a function of field for a Pt(2.5nm)/Ru(2nm)/Fe$_3$Sn$_2$(80nm) film (b) and a 40 nm Py (c) film. Inset of (b): $E_{yx}$ (left $y$-axis) and $\nabla T_x$ (right $y$-axis) as a function of heating power. (d), (e) $E_{yz}$ ($\nabla T_z$, $H \parallel ab$-plane) as a function of field for Fe$_3$Sn$_2$ (d) and Py (e) films.